\documentclass[12pt]{iopart}
\pdfoutput=1
\usepackage{iopams}

\usepackage{amsfonts}
\usepackage{cite}
\usepackage{graphicx}
\usepackage{epstopdf} 
\usepackage{amsthm}
\usepackage{color}
\usepackage[caption=false]{subfig}

\usepackage{verbatim}

\usepackage{array}

\usepackage{kbordermatrix}

\begin{document}

\title{Improving power grid transient stability by plug-in electric vehicles}

\author{Andrej Gajduk$^1$, Mirko Todorovski$^2$, Juergen Kurths$^{3,4,5}$,  and Ljupco Kocarev$^{1,6,7}$\footnote{Author to whom any correspondence should be addressed}}

\address{$^1$ Research Center for Computer Science and Information Technologies, Macedonian Academy of Sceinces and Arts, Skopje, Macedonia\\
$^2$ Faculty of Electrical Engineering and Information Technologies, ``Ss Cyril and Methodius'' University, Skopje, Macedonia\\
$^3$ Potsdam Institute for Climate Impact Research, PO Box 60 12 03, Potsdam 14412, Germany \\
$^4$ Department of Physics, Humboldt University of Berlin, Newtonstrasse 15, Berlin 12489, Germany \\
$^5$ Institute for Complex Systems and Mathematical Biology, University of Aberdeen, Aberdeen AB24 3UE, UK \\
$^6$ Faculty of Computer Science and Engineering, ``Ss Cyril and Methodius'' University, Skopje, Macedonia\\
$^7$ BioCircuits Institute, UC San Diego, La Jolla, CA 92093-0402, USA}
\ead{lkocarev@ucsd.edu}
\vspace{10pt}
\begin{indented}
\item[]July 2014
\end{indented}

\begin{abstract}
Plug-in electric vehicles (PEVs) can serve in discharge mode as distributed energy and power resources operating as vehicle-to-grid (V2G) devices and in charge mode as loads or grid-to-vehicle (G2V) devices. It has been documented that PEVs serving as V2G systems can offer possible backup for renewable power sources, can provide reactive power support, active power regulation, load balancing, peak load shaving,
can reduce utility operating costs and can generate revenue. Here we show that PEVs can even improve power grid transient stability, that is, stability when the power grid is subjected to large disturbances, including bus faults, generator and branch tripping, and sudden large load changes. A control strategy that regulates the power output of a fleet of PEVs based on the speed of generator turbines is proposed and tested on the New England 10-unit 39-bus power system. By regulating the power output of the PEVs we show that (1) speed and voltage fluctuations resulting from large disturbances can be significantly reduced up to 5 times, and (2) the critical clearing time can be extended by 20-40\%. Overall, the PEVs control strategy makes the power grid more robust.
\end{abstract}

%
%
\submitto{\NJP}
%
\maketitle
%
%

\tableofcontents

\section{Introduction}

During the last century, two large and separate systems for energy conversion and management were developed: the electric utility system and the light vehicle fleet. The electric utility system converts stored energy (chemical, mechanical, and nuclear) to electric energy and delivers it to customers through interconnected transmission and distribution grids. The light vehicle fleet (passenger cars, vans, and light trucks) converts petrochemical energy to kinetic energy (rotary motion) for providing moving/traveling of citizens and goods. 

The electric power grid and light vehicle fleet are exceptionally complementary as systems for converting and managing energy and power \cite{kempton2005vehicle1,kempton2005vehicle2}. According to \cite{kempton2005vehicle1,kempton2005vehicle2}, in 2005 the total power capacity of the US light vehicles fleet was 24 times the power capacity of the US entire electric generation system. The electric utility system has essentially no storage, and therefore, energy/power generation and transmission must be continuously managed to match fluctuating customer load. The light vehicle fleet, however, inherently has storage, since both the vehicle and its fuel must be mobile. The high capital cost of large generators motivates high use (average 57\% capacity factor). On the other hand, personal vehicles are cheap per unit of power and are utilized only 4\% of the time for transportation, making them potentially available the remaining 96\% of time for a secondary function~\cite{kempton2005vehicle1,kempton2005vehicle2}.

Recently, there has been increased concern on environmental and climate change issues, rising energy costs, energy security and fossil energy reserves, which, in turn, has triggered worldwide interest for plug-in electric vehicles (PEVs)~\cite{ehsani2009modern,larminie2003electric,wirasingha2009pihef}. Plug-in vehicles can behave either as loads, which is usually referred as grid-to-vehicle (G2V) connection, or as a distributed energy and power resource in a concept known as vehicle-to-grid (V2G) connection. PEVs have not yet been widely adopted, in part because of technical limitations, social obstacles, and cost compared to conventional internal combustion engine vehicles~\cite{sovacool2009beyond}. However, based on moderate expectations, by 2020 up to 2\% of the total vehicles in the U.S. will be PEVs according to the Oak Ridge National Laboratory (ORNL)~\cite{sikes2010plug} or 1\% according to more conservative expectations~\cite{duan2014forecasting}.  Moreover, there is enough generation capacity in the U.S. to absorb one million or more PEVs without shortage~\cite{rahman2012mitigation}.	

PEVs can serve in discharge mode as vehicle-to-grid (V2G) devices and in charge mode as grid-to-vehicle (G2V) devices~\cite{kempton2005vehicle1,kempton2005vehicle2}. Economic costs, emissions benefits, and distribution system impacts of PEVs depend on vehicle and battery characteristics as well as on charging and recharging frequency and strategies.
Previous studies have shown that smart charging, which can be done by means of smart metering, control, and communication, minimizes PEV impact on the power grid~\cite{yilmaz2013review,qian2011modeling,duvall2007environmental,sioshansi2009emissions,samaras2008life,de2006grid}. 
A strategy for reducing operational cost and
emission for grid operators is suggested in~\cite{saber2010intelligent} and consists of offering real-time electricity pricing for charging and discharging. Each vehicle can be contracted individually or as part of an aggregation. Aggregators collect PEVs into a group to create a larger, more manageable load for the utility~\cite{guille2009conceptual}. These groups can act as distributed energy resources to realize ancillary services and spinning reserves.
Cooperation between the grid operator and vehicle owners or aggregators is essential in integrating both systems.

Potential benefits, including costs issues, of the V2G concept have been very active research topics [14--29]. V2G systems: (1) can offer a possible backup for renewable power sources including wind and solar power, supporting efficient integration of intermittent power production~\cite{marano2008energy,short2006preliminary,ramos2008modeling}; (2) can provide reactive power support~\cite{de2006grid}, active power regulation, load balancing by valley filling~\cite{koyanagi1998strategy,takagi2012electricity}, peak load shaving~\cite{kintner2007impacts,sortomme2012optimal}, and current harmonic filtering; (3) can provide ancillary services as frequency control and spinning reserves~\cite{guille2009conceptual,kempton2001vehicle,wirasingha2008plug,dallinger2011vehicle,keane2012potential} ; (4) can improve grid efficiency, stability, reliability~\cite{srivastava2010challenges}, and generation dispatch~\cite{denholm2006evaluation}; and can reduce utility operating costs and even potentially can generate revenue~\cite{saber2010intelligent}. 
%

Depending upon the nature of disturbance in power system, the stability problems are classified into two categories: steady state stability (small signal stability) and transient stability.
A power system is steady state stable if it is able to reach a new stable configuration following a small disturbance in the system~\cite{sauer1990power}. 
The new stable state is very close to the pre-disturbance operating point. In such cases the equations describing the power system dynamics may be linearized for analytical purposes. Steady state stability depends on the initial operating point of the system and it is closely connected to the voltage regulation policy, as well as loading of lines/transformers.
%
Power systems may experience severe disturbances such as: short circuits with or without significant network topology change, switching operations of important lines/transformers, and sudden demand or insufficiency of load~\cite{dash2000classification}. Transient stability of a power system refers to the ability of the system to reach a stable condition following a large disturbance in the transmission network. 
Power grid stability has attracted recently interest also by physics community, see for example \cite{rohden2012self,rohden2013impact,dorfler2013synchronization,menck2013basin,cornelius2013realistic}. An overview of the progress in transient and steady-state stability of power grids is provided in~\cite{gajdukstability}.


In this paper we show that PEVs can improve power grid transient stability, that is, stability when the power grid is subjected to large disturbances, which are characterized by bus faults, generator and branch tripping, and sudden large load changes. 
A control strategy that regulates the power output of a fleet of PEVs based on the speed of generator turbines is proposed in section 2. This section also describes the power system model: we propose a hybrid approach that eliminates only buses that do not include aggregated V2G systems. For testing the suggested control strategy a simulation tool PSS/E, developed by Siemens, is used. Section 3 summarizes the results and discussions. The control strategy is tested on the New England 10-unit 39-bus power system~\cite{pai1989energy}. By regulating the power output of the PEVs we show that (1) the speed and the voltage fluctuations followed a large disturbance can be significantly reduced up to 80\%, and (2) the critical clearing time can be extended by 20-40\%. In section 4, we conclude our paper.  

\section{Materials and Methods}

In power grids, automatic generation control (AGC) aims to balance generation and load, by adjusting the power output of the generators.
Typical response times for AGC are in the order of minutes.
This type of generation control is not suited to tackle large disturbances where required reaction times are much shorter, usually just several seconds~\cite{machowski2011power}.
%
We propose a novel control strategy that regulates power exchange between PEVs and the power grid, based on the turbine speeds at the generators, in an effort to reduce the effects of large disturbances.
A decrease in the turbine speeds signals a power shortage, thus PEVs are instructed to feed additional power to the grid, in essence acting as small generators.
On the other hand, when there is surplus power in the grid, indicated by increased turbine speeds, the PEVs act as loads and consume any excess power.
Note that the PEVs only exchange real power because that is the quantity related to the turbine speeds.

Effective integration of PEVs for regulatory purposes requires that information from the generators reaches individual vehicles within 10--20 milliseconds, according to IEEE standard 1646.
To meet these strict requirements, scientists suggest using a dedicated, high-speed and high-bandwidth wirelane cables, \textit{e.g.} 1 Gbps Ethernet~\cite{xiaoyang2005analysis}.
Furthermore, it is proposed to use Wavelength Division Multiplexing (WDM) to increase the bandwidth and advanced protocols such as MultiProtocol Label Switching (MPLS) to reduce delays.
Besides latency, which is the most critical performance metric, the communication infrastructure must satisfy high-requirements for reliability and security within the delay constraint.
For a more detailed analysis on the communication requirements for application in smart grid we refer readers to~\cite{cho2006overview,wang2011survey}.

%

Regulating the power grid requires PEVs to provide additional power, which reduces the amount of energy available for driving.
Fortunately, large disturbances which are the focus of our work, occur rarely, several hundred times per year~\cite{grigg1999ieee,kjole2011stats}. 
Furthermore, the regulatory actions that follow large disturbances usually last only a few seconds.
Consequently, the effects of power regulation on the vehicle's energy reserves is minimal~\cite{kempton2005vehicle1,kempton2005vehicle2}.
Additionally, PEVs are parked and plugged-in 96\% of the time, making them available for regulatory actions~\cite{santos2011summary}.


To facilitate control, PEVs connected to the same substation are aggregated together to form an \textit{automatically controlled vehicle group} (ACVG).
The ACVGs are the basic control units used in our power system model.

\subsection{Power system model}

In order to asses the ability of PEVs to regulate the power grid, we develop a power system model that incorporates ACVGs into the grid architecture.
In our model, ACVGs are treated as constant power loads that can take positive as well as negative values.
This means that they can both absorb power from the grid and also inject power to the grid.
Please note that the term constant power refers to the fact that the power exchanged between the vehicle and grid is neither frequency nor voltage dependent.
Such modeling is realistic, since PEVs are connected to the grid via power electronics which can regulate the power exchange regardless of the voltage and frequency levels in the grid. The previous statement may not completely hold for a very short period of time ($\approx 0.1~{\rm sec}$) when a short circuit is present, since in that case severe voltage drops are possible in the vicinity of the short circuit location and PEVs in that area may not completely behave as constant power sources.

There are two main approaches to power system modeling.
The first approach is designed to enable engineers to focus only on the generator buses, which drastically reduces the problem size. This is justified by the main engineering concern  whether all generators will remain in synchronous operation or not. 
In this case, all loads are modeled with constant admittances and the buses where they are connected are eliminated via Kron reduction.
This model is also known as the \textit{classical model}~\cite{kundur1994power,filatrella2008analysis,machowski2011power}.
As ACVGs are modeled as constant power loads this approach is not applicable.

The structure preserving model is a different approach that is used to study power systems.
As its name implies it preserves the structure \textit{i.e.} the topology of the grid.
This approach supports constant power loads. However, since all the buses are retained, this model is not suited for large power systems where only few of the buses include an ACVG.
Therefore we propose a hybrid approach that eliminates only those buses that do not include a generator or an ACVG.
A detailed description of this approach follows.

Consider a power system that consist of $N$ buses of which: $n$ generator buses numbered $1,\ldots,n$; $m$ ACVG buses numbered $n+1,\ldots,n+m$ and $N-(n+m)$ stub buses numbered $n+m+1,\ldots,N$.
Any bus can also include one or more classical loads.
Generators are modeled as a constant electromotive force behind a transient reactanse.
Two types of load models are used: classical loads are represented as constant admittances, while ACVGs are represented as constant power loads.

The bus admittance matrix $\textbf{Y}_{bus}$ is obtained with the incidence matrix, which holds information on branch-to-node connections, and the primitive admittance matrix formed from $\pi$-equivalent circuits for all network elements~\cite{grainger1994power}. 
Classical loads are directly included into the admittance matrix, resulting in
\begin{equation}
\label{eq:admittance_matrix}
  \textbf{Y}_{bus}=\kbordermatrix{%
        & n   & m & N-(n+m) \\
    n   & \textbf{Y}_{G,G}+\textbf{Y}_{l,G} & \textbf{Y}_{G,A} & \textbf{Y}_{G,S} \\
   m & \textbf{Y}_{G,A}^{\rm T} & \textbf{Y}_{A,A}+\textbf{Y}_{l,A} & \textbf{Y}_{A,S} \\
    N-(n+m) & \textbf{Y}_{G,S}^{\rm T} & \textbf{Y}_{A,S}^{\rm T} & \textbf{Y}_{S,S}+\textbf{Y}_{l,S}
  }
\end{equation}
where $G, A$ and $S$ stands for generator buses, ACVG buses, and stub buses, respectively. The admittance matrix  $\textbf{Y}_{G,A}$ is a mutual admittance matrix between generator and ACVG buses. Other notations follow by analogy. The  constant admittance loads at the generator, ACVG, and stub buses are denoted by $\textbf{Y}_{l,G}$, $\textbf{Y}_{l,A}$ and $\textbf{Y}_{l,S}$, respectively.

Next, the admittance matrix is expanded with $n$ additional buses that represent the fictious generators.
These buses are inserted at positions $1,\ldots ,n$ and are connected to buses $n+1,\ldots ,2n$ by admittances that correspond to the generators transient reactanses 
\begin{equation*}
\textbf{Y} = Diag(Y_{gi}), \mbox{ where }  Y_{gi} = \frac{1}{jX_{gi}}
\end{equation*}
The augmented bus admittance matrix thus becomes
\begin{equation*}
  \hat{\textbf{Y}}_{bus}=\kbordermatrix{%
        & n & n  & m & N-(n+m) \\
    n   & \textbf{Y}  & -\textbf{Y} &  \textbf{0} & \textbf{0} \\
    n   & -\textbf{Y} & \textbf{Y}+\textbf{Y}_{G,G}+\textbf{Y}_{l,G} & \textbf{Y}_{G,A} & \textbf{Y}_{G,S} \\
   m & \textbf{0} &  \textbf{Y}_{G,A}^{\rm T} & \textbf{Y}_{A,A}+\textbf{Y}_{l,A} & \textbf{Y}_{A,S} \\
    N-(n+m) & \textbf{0} & \textbf{Y}_{G,S}^{\rm T} & \textbf{Y}_{A,S}^{\rm T} & \textbf{Y}_{S,S}+\textbf{Y}_{l,S}
  }
\end{equation*}
Buses that do not include generators or ACVGs can be safely removed without any loss of information, in order to reduce the size of the admittance matrix.
First, the admittance matrix needs to be rearranged into four block matrices 
\begin{equation*}
  \hat{\textbf{Y}}_{bus}=\kbordermatrix{%
        & n & m &  & n & N-(n+m) \\
    n   & \textbf{Y}  & \textbf{0} & \vrule &  -\textbf{Y} & \textbf{0} \\
    m & \textbf{0} & \textbf{Y}_{A,A}+\textbf{Y}_{l,A}\hspace{-3pt} & \vrule & \textbf{Y}_{G,A}^{\rm T}  & \textbf{Y}_{A,S} \\[1.8ex]
    \cline{2-6}
 & & &\vrule&  & \\
    n   & -\textbf{Y} & \textbf{Y}_{G,A} & \vrule & \hspace{-5pt} \textbf{Y}+\textbf{Y}_{G,G}+\textbf{Y}_{l,G}  & \textbf{Y}_{G,S} \\
    N-(n+m) & \textbf{0} & \textbf{Y}_{A,S}^{\rm T} & \vrule & \textbf{Y}_{G,S}^{\rm T}  & \textbf{Y}_{S,S}+\textbf{Y}_{l,S}
  }
\end{equation*}
\begin{equation*}
  \hat{\textbf{Y}}_{{bus}} =
  \kbordermatrix{%
        & n+m   &  N-m  \\
    n+m   & \textbf{Y}_a & \textbf{Y}_b \\
	N-m   & \textbf{Y}_b^T & \textbf{Y}_c \\
  }
\end{equation*}
The reduced admittance matrix can then be derived by applying Kron reduction~\cite{dorfler2013kron}
\begin{equation*}
\textbf{Y}_{red} = \textbf{Y}_a-\textbf{Y}_b\textbf{Y}_c^{-1}\textbf{Y}_b^T
\end{equation*}
Each element of $\textbf{Y}_{red}$ is comprised of a real and an imaginary part $Y_{ik} = G_{ik}+\jmath B_{ik}$.

At this moment, let us adopt the following notation: $\phi_i$ is the bus phase angle at the $i$th bus with respect to an arbitrary reference bus, while $V_i$ is the voltage at bus $i$. The mechanical dynamics of the $i$th generator are governed by the swing equation
\begin{equation}
\label{eq:swing}
M_i \ddot{\phi_i} + D_i \dot{\phi_i} = P_{mi} - P_{ei}
\end{equation}
where $M_i$ is the time inertia constant of the rotor, $D_i$ is the damping constant of the rotor, $P_{mi}$ is the mechanical output power, and $P_{ei}$ is the electrical power of the $i$th generator.

The electrical power is the real part of the complex power at bus $i$ 
\begin{equation}
\label{eq:elec_power}
P_{ei} = \Re(S_i) = \Re(V_iI_i^*)
\end{equation}
The complex current can be expressed in terms of the voltage and admittance as 
\begin{equation}
\label{eq:complex_current}
I_i = \sum_{k=1}^{n+m} Y_{ik}V_k
\end{equation}
Substituting Eq.~(\ref{eq:complex_current}) into Eq.~(\ref{eq:elec_power}) and applying the Euler's formula results in
\begin{equation}
\label{eq:elec_power_full}
P_{ei} = V_i\sum_{k=1}^{n+m}V_k\left[ G_{ik} \cos (\phi_i-\phi_k)+B_{ik} \sin (\phi_i-\phi_k) \right]
\end{equation}
The effects of the turbine governors can be neglected due the short time scales involved with transient stability. The mechanical power is thus constant, and equal to the electrical power at the moment before the fault occurs.

By introducing an additional variable $\omega_i = \dot\phi_i$ we can reduce Eq.~(\ref{eq:swing}) to a set of first order differential equation. Finally, by substituting Eq.~(\ref{eq:elec_power_full}) the following set of ODEs is derived
\begin{eqnarray}
\label{eq:swing_full}
  \dot \phi_i  &=& \omega_i \nonumber \\
  \dot \omega_i  &=& \frac{1}{M_i} \left[ P_{mi} - D_i\omega_i - V_i\sum_{k=1}^{n+m}V_k\left[ G_{ik} \cos (\phi_{ik})+B_{ik} \sin (\phi_{ik}) \right]\right]
\end{eqnarray}
where $i =1,\ldots,n$ and $\phi_{ik} = \phi_i-\phi_k$.

In addition to this set of $2n$ differential equations that describe the mechanical dynamics of the generators, $2m$ algebraic equations are needed to describe the voltage-power relation of the ACVGs.
As mention previously, the ACVGs are modeled as constant power loads with real power $P_{i}^{ACVG}$ and no reactive power.
The equations for real and reactive power at the $i$th ACVG bus, $i=n+1,\ldots,n+m$, are
\begin{eqnarray}
  P_i^{ACVG} + V_i \sum_{k=1}^{n+m} V_k \left( G_{ik}  \cos (\phi_i - \phi_k)+ B_{ik} \sin (\phi_i - \phi_k) \right) &=& 0 \nonumber \\
  V_i\sum_{k=1}^{n+m} V_k \left(G_{ik}   \sin (\phi_i - \phi_k) - B_{ik} \cos (\phi_i - \phi_k) \right) &=& 0,  \label{eq:power_acvg_full}
\end{eqnarray}
Combining the swing equations (\ref{eq:swing_full}) with Eqs.~(\ref{eq:power_acvg_full}) leads to a set of differential algebraic equations (DAE), that fully describe the dynamics of the power system.

In general each of the algebraic equations has $n + m$ voltage terms out of which $n$ are already known being the constant electromotive forces of the generators.
This leads to $m$ non-linear equations in $m$ unknown variables which can be numerically solved by standard methods such as the Newton-Raphson method.
Once the voltages on both the generator and ACVG buses are known, the bus current injection and the electrical power output of the $i$th machine can be calculated using Eqs.~(\ref{eq:complex_current}) and (\ref{eq:elec_power}).

\subsection{Control strategy}

In order to stabilize the system, the power of the $i$th ACVG, $P_i^{ACVG}$, is set to be a function of the turbine speeds
\begin{equation}
\label{eq:ACVG_power2}
P_i^{ACVG} = \Delta \omega h_i 
\end{equation}
where 
$\Delta \omega = \frac{1}{n}\sum_{k=1}^n \omega_k-\omega_{ref}$ is the average frequency deviation, and $h_i$ is a control coefficient. 
This is a simple linear control function that is similar to automatic generation control (AGC) in turbines. However, unlike AGC which only considers the frequency at the local generator, our type of control takes into account information from all the generators.

%
%

Various limitations of PEVs must be considered, in order to identify the correct value for the control coefficient at each ACVG bus. The current-carrying capacity of the connecting infrastructure is the main limiting factor in the vechicle-to-grid implementation~\cite{kempton2005vehicle1,kempton2006electric}. 
We take that a single plug-in vehicle can exchange electricity at a maximum rate of 10 kW, which is a rather conservative number~\cite{morrow2008plug}. Larger exchange rates would allow for more regulating power, thus enabling the system to deal with more massive disturbances. If the total number of PEVs in the power system is $N_{PEV}$ then the maximal power exchange between the grid and the combined fleet of PEVs is
\begin{equation}
\label{eq:max_total_power}
P_{max} = 10 N_{PEV} \mbox{[kW]}
\end{equation}

In general, the PEVs can be distributed arbitrarily around the power system. However, for large power systems that serve millions of customers, one expects little variation in the number of PEVs in different areas with respect to their population, because peoples' habits and lifestyle would be similar.
Thus, we assume that PEVs are distributed uniformly throughout the power system, and that the number of PEVs at a particular ACVG bus is proportional to the energy consumption at that bus.
This can be justified by the fact that energy consumption is indicative of the population size. 
Namely, the residential and commercial sector, which constitute 33.6\% and 32.4\% of the total energy consumption in the US, respectively, are directly correlated with population, since there are only small discrepancies in the average household consumption.
Under this assumption, the bounds for power exchange between the $i$th ACVG bus and the grid are determined by
\begin{equation}
-P_{max} \frac{P_i}{\sum_{k=1}^N P_k} [\mbox{kW}] \leq  P_{i}^{ACVG} \leq P_{max} \frac{P_i}{\sum_{k=1}^N P_k} [\mbox{kW}]
\end{equation}

If the distribution of PEVs is not uniform the control strategy should be adjusted to account for that. Currently, in such cases the control may overload certain transmission lines by injecting too much power from a PEV rich area. However, this overload lasts several seconds and it should not cause problems since the line protection acts with time delay which is much longer. Only in extreme cases, with overload being 2-3 times bigger than the normal load, line tripping may occur in matter of seconds or even a fraction of a second.

In most situations a speed deviation in the grid needs to be met with generation and load changes as soon as possible, but without overcompensating. In the proposed control strategy we aim to achieve the maximal power exchange between the ACVGs and the grid when the average speed deviation reaches $\pm 100$ mHz. The control coefficient for the $i$th ACVG can then be calculated using
\begin{equation}
h_i = \frac{P_{j,max}^{ACVG}}{0.1} =  100 N_{PEV} \frac{P_i}{\sum_{k=1}^N P_k} \left[ \frac{\mbox{kW}}{\mbox{Hz}}\right]
\end{equation}
If the average speed deviation exceeds $\pm 100$ mHz, the power of the ACVGs remains at the same maximum level. The power of the $i$th ACVG bus is

$$
P_i^{ACVG} = \left\{ \begin{array}{rl}
-10 N_{PEV} \hat{P}_i \hspace{8pt} \mbox{[kW],} &\mbox{ if $\Delta \omega \leq -100$ [mHz]} \\
100 N_{PEV} \hat{P}_i \Delta \omega \hspace{8pt} \mbox{[kW],} &\mbox{ if $-100 < \Delta \omega \leq 100 $  [mHz]} \\
10 N_{PEV} \hat{P}_i \hspace{8pt} \mbox{[kW],}&\mbox{ if $\Delta \omega > 100$ [mHz]}
\end{array} \right.
$$
where $i = 1,\ldots,m$ and $\hat{P}_i = \frac{P_i}{\sum_{k=1}^N P_k}$.

\subsection{Simulations method}
\label{sec:simulations_method}

We tested the proposed control strategy on the New England power system, which is often used when examining power grid stability~\cite{llamas1995clarifications}.
The decision to use the New England power system was arbitrary, mainly because it is easily accessible, widely used in literature, and it models part of a real power system.
This system is comprised of 39 buses, 10 generation units and 17 loads all of which are connected by 48 transmission lines.
We modified the original test set by adding 17 ACVGs at each bus that has at least one load connected to it.
Initially, the ACVGs power is set to $0$ and any changes are due to automatic control actions.
The generator connected to bus $31$ is used as the swing machine.
No turbine governor models are used for the generators.

In order to calculate the control coefficients $h_i$ for the ACVG, we need to estimate the number of PEVs in New England.
At the time the New England power system is proposed as a test system, the total population in New England was around 10 million citizens.
According to USA statistics of the time there was, on average, one car for every two citizens, which would sum up to a total of 5 million passenger cars.
Assuming 1\% PEV penetration (which should happen by 2020) the number of PEVs available for power grid regulation is $N_{PEV} = 50\hspace{3pt}000$. 

For efficient dynamic simulations of power systems we use the well-known software tool PSS/E, developed by Siemens.
This tool supports the 2 load types used in our model i.e. constant power and constant admittance.
Additionally, the output power of ACVGs can be changed during simulations and can receive both positive and negative values.
The software tool is able to simulate a broad range of disturbances including: bus faults, line tripping and sudden load changes, which are used in our simulations.

\section{Results and Discussion}

\begin{figure}[!htb]
\centering
\begin{tabular}{ c  c  c }
\captionsetup[subfigure]{margin={0.7cm,0cm}}
\subfloat[]{\includegraphics[scale=.19]{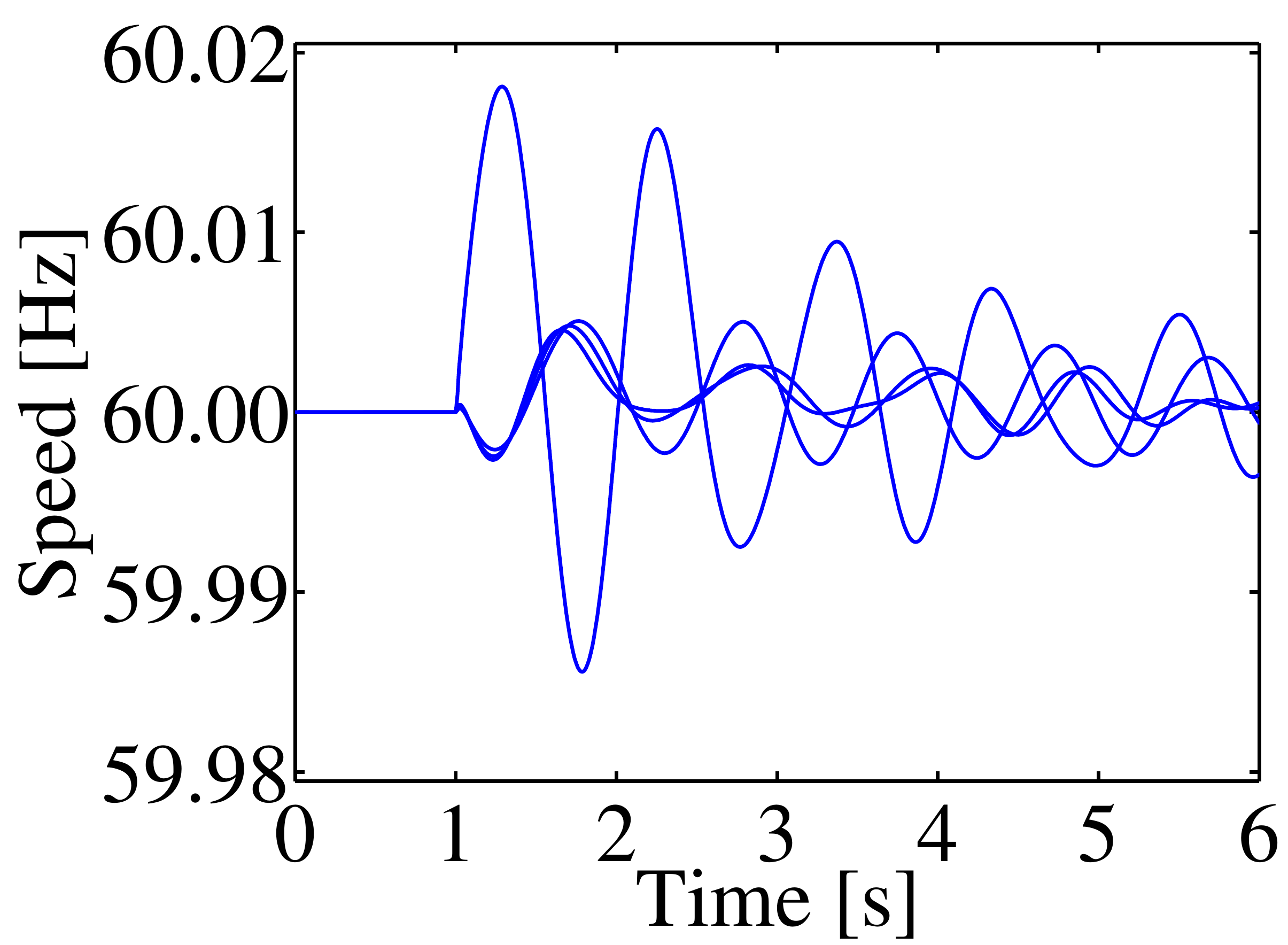}}
&
\captionsetup[subfigure]{margin={0.7cm,0cm}}
\subfloat[]{\includegraphics[scale=.19]{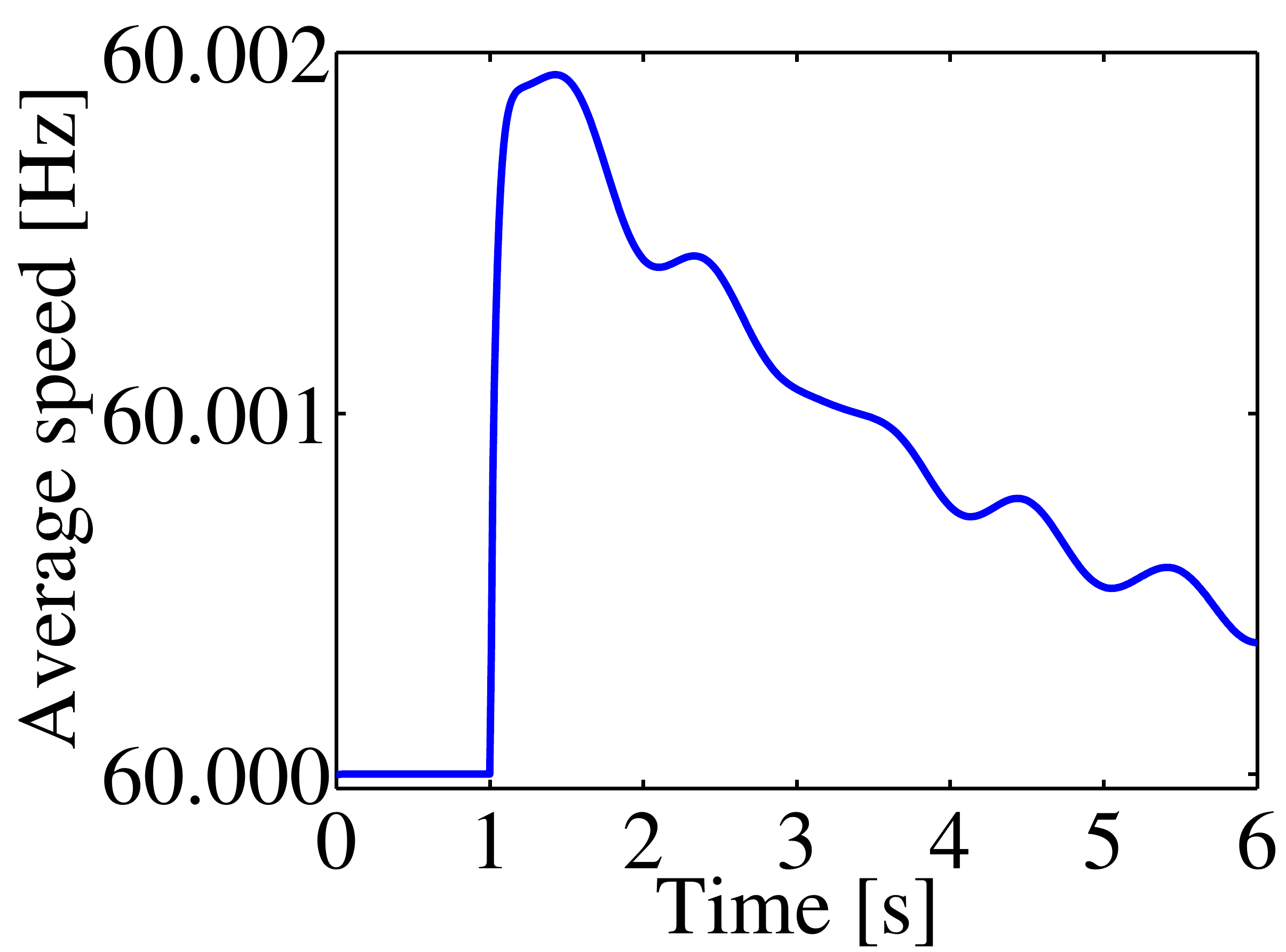}}
&
\captionsetup[subfigure]{margin={0.6cm,0cm}}
\subfloat[]{\includegraphics[scale=.19]{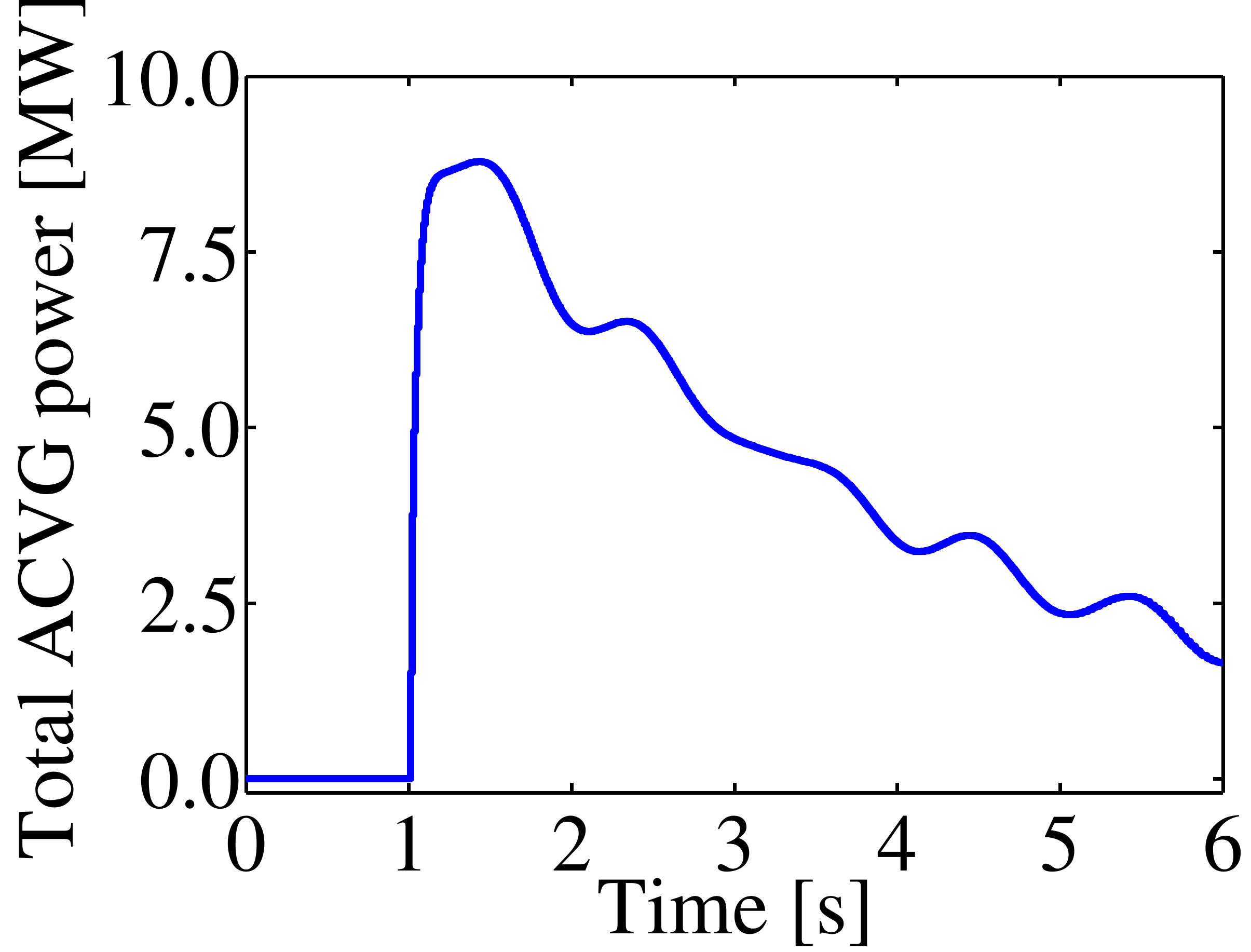}}
\end{tabular}
\caption{Changes in the turbine speeds due to a sudden load decrease of 10\% at bus 20 which starts at the 1 second mark (a). As expected, the lack of load leads to a surplus in power which results in increased average speed (b). This change is quickly met by the ACVGs which act as loads in order to absorb the excess power (c).}  
\label{fig:load_change}
\end{figure}

As a demonstration of the ACVGs power regulation in response to changes in the generator's turbine speeds, we simulated a 10 percent abrupt load decrease at bus 20.
The decrease occurs at the 1 second mark and lasts for the entire length of the simulation.
The resulting speed deviations, together with the ACVGs power are shown in Fig.~\ref{fig:load_change}. 
Immediately after the load decreases the system frequencies begin to deviate from their nominal value.
The ACVGs react promptly by assuming the role of loads.
After several seconds the system begins to stabilize and involvement of ACVGs subsides.

When large disturbances such as branch faults and line trips occur, the power system is subjected to speed and voltage fluctuations.
These fluctuations affect the power quality that customers experience, and if they are strong enough they could even damage sensitive appliances.
Moreover, if the disturbance is not cleared within a specific time, called the critical clearing time, then the system will be unable to resume stable operation.

Interestingly, by integrating PEVs into the power grid
both these negative effects are significantly  lessened.
Firstly, the speed and voltage fluctuations that follow a non-critical fault are reduced by a factor of 2-6.
Secondly, the critical clearing times are extended by 20-40\%.

\begin{figure}[!htb]
\centering
\begin{tabular}{ c  c}
\captionsetup[subfigure]{margin={0.7cm,0cm}}
\subfloat[]{\includegraphics[scale=.2]{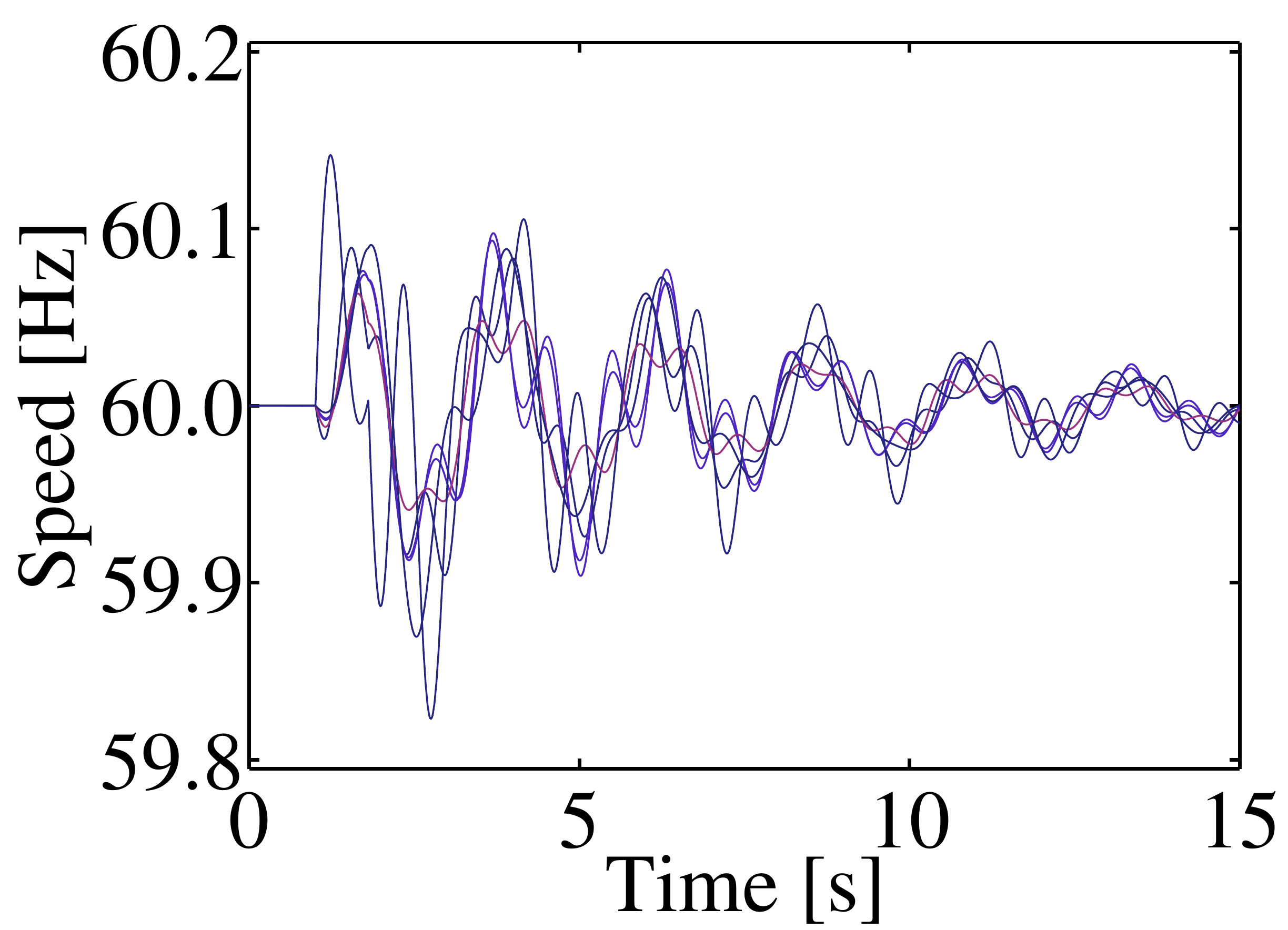}}
&
\captionsetup[subfigure]{margin={0.7cm,0cm}}
\subfloat[]{\includegraphics[scale=.2]{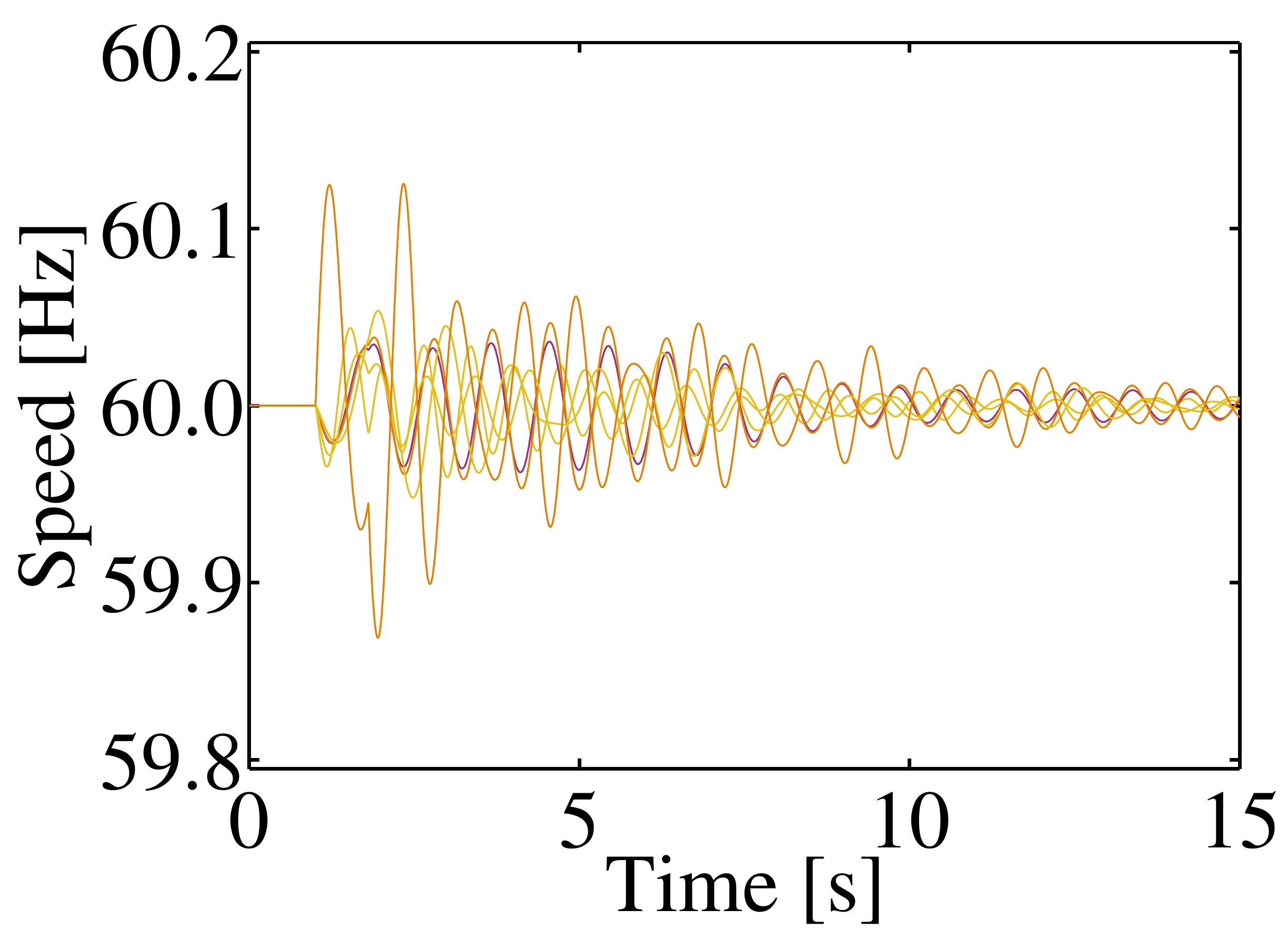}}
\end{tabular}
\caption{Speed fluctuations following a line trip between buses 23 and 24, that starts at the 1 second mark and lasts for 0.1 seconds. The lines are colored according to the magnitude by which the deviate from the nominal value, with purple(dark grey) corresponding to large deviations, while yellow(light grey) indicates little or no deviation. Without ACVGs speeds strongly deviate from their nominal values and take a long time to stabilize as indicated by dominantly purple color (a). On the other hand, when ACVGs are used, these fluctuations are considerably reduced (b).} 
\label{fig:reduced_fluctuations}
\end{figure}

We compared the speed and voltage fluctuations between the original New England test system and the New England with added ACVGs by simulating different types of disturbances such as bus faults, generators trips, and sudden large load changes.
The results for a sample of simulated disturbances show that speed and voltage fluctuations are reduced by 50-80\% when PEVs are used as a control mechanism (Table~\ref{tab:red_fluc2}).
An example that illustrates the typical behavior of the system when a disturbance occurs is given in Fig.~\ref{fig:reduced_fluctuations}.
Please note that we tested many different disturbance scenarios, all of which produced similar results.

The fluctuation reductions are lowest for branch trips, as opposed to other types of disturbances (Table~\ref{tab:red_fluc2}).
A possible explanation is that during a branch trip the system ``splits'' into two subsystems, one subsystem has excess power, while the other subsystem has a shortage of power.
In such cases it could be better to control PEVs with regards to local conditions, instead of global ones.

\begin{table}[!b]
\caption{Reduction in speed and voltage fluctuation for several disturbances. The deviations were calculated in terms of the square difference from their respective nominal values, then averaged across all buses.}
\label{tab:red_fluc2}
\begin{center}
\begin{tabular}{ | c | c | c | c | c | c | c | c |  }
\hline
 \multicolumn{2}{|c|}{} & \multicolumn{3}{c}{speed deviation} & \multicolumn{3}{|c|}{voltage deviation} \\ \hline
Disturbance & t [s] & NO PEVs & PEVs & \% &  NO PEVs & PEVs & \% \\ \hline
bus fault 6 & 0.07 & 0.6012 & 0.1025 & 83 & 8.8986 & 2.9690 & 67 \\ \hline
bus fault 28 & 0.07 & 0.3883 & 0.1254 & 68 & 3.8113 & 1.9265 & 49 \\ \hline
branch trip 3-18 & 0.15 & 2.39$\cdot 10^{-3}$ & 8.78$\cdot 10^{-4}$ & 63 & 0.0035 & 0.0008 & 77 \\ \hline
branch trip 16-21 & 0.15 & 0.0884 & 0.0415 & 53 & 0.1192 & 0.0425 & 64 \\ \hline
load decrease 20\% 7 & 5.00 & 4.27$\cdot 10^{-4}$ & 9.01$\cdot 10^{-5}$ & 79 & 0.0155 & 0.0019 & 87 \\ \hline
load increase 20\% 29 & 5.00 & 4.00$\cdot 10^{-3}$ &  1.35$\cdot 10^{-3}$ & 66 & 0.2061 & 0.1014 & 51 \\ \hline
\end{tabular}
\end{center}
\end{table}

The increase in the critical clearing times for several sample bus faults is summarized in Table~\ref{tab:acvg_t_ccl}.
This increase is consistent for all types of buses: generator buses (32,37); load buses (4,16,20,24); and stub buses (1,11).
Bus faults are the only type of disturbance considered since other types like load changes are less severe and can not lead to a loss of stability.

\begin{table}[!htb]
\caption{A comparison between the critical clearing times with and without ACVG.}
\label{tab:acvg_t_ccl}
\begin{center}
\begin{tabular}{ | c || c | c | c | c | c | c | c | c | }
\hline
 Bus               &  1       &  4       &  11      &  16      &  20      &  24      &  32      &  37         \\ \hline
 $t_{ccl}$         &  0.2732  &  0.1101  &  0.1210  &  0.0844  &  0.1189  &  0.0899  &  0.1508  &  0.1705     \\ \hline
 $t_{ccl}^{ACVG}$  &  0.3850  &  0.1396  &  0.1503  &  0.1000  &  0.1520  &  0.1242  &  0.1848  &  0.2111     \\ \hline
\% increase        &  40.922  &  26.793  &  24.214  &  18.483  &  27.889  &  38.153  &  22.546  &  19.232     \\ \hline
\end{tabular}
\end{center}
\end{table}

\begin{figure}[!htb]
\centering
\begin{tabular}{ c  c }
\captionsetup[subfigure]{margin={0.7cm,0cm}}
\subfloat[]{\includegraphics[scale=.2]{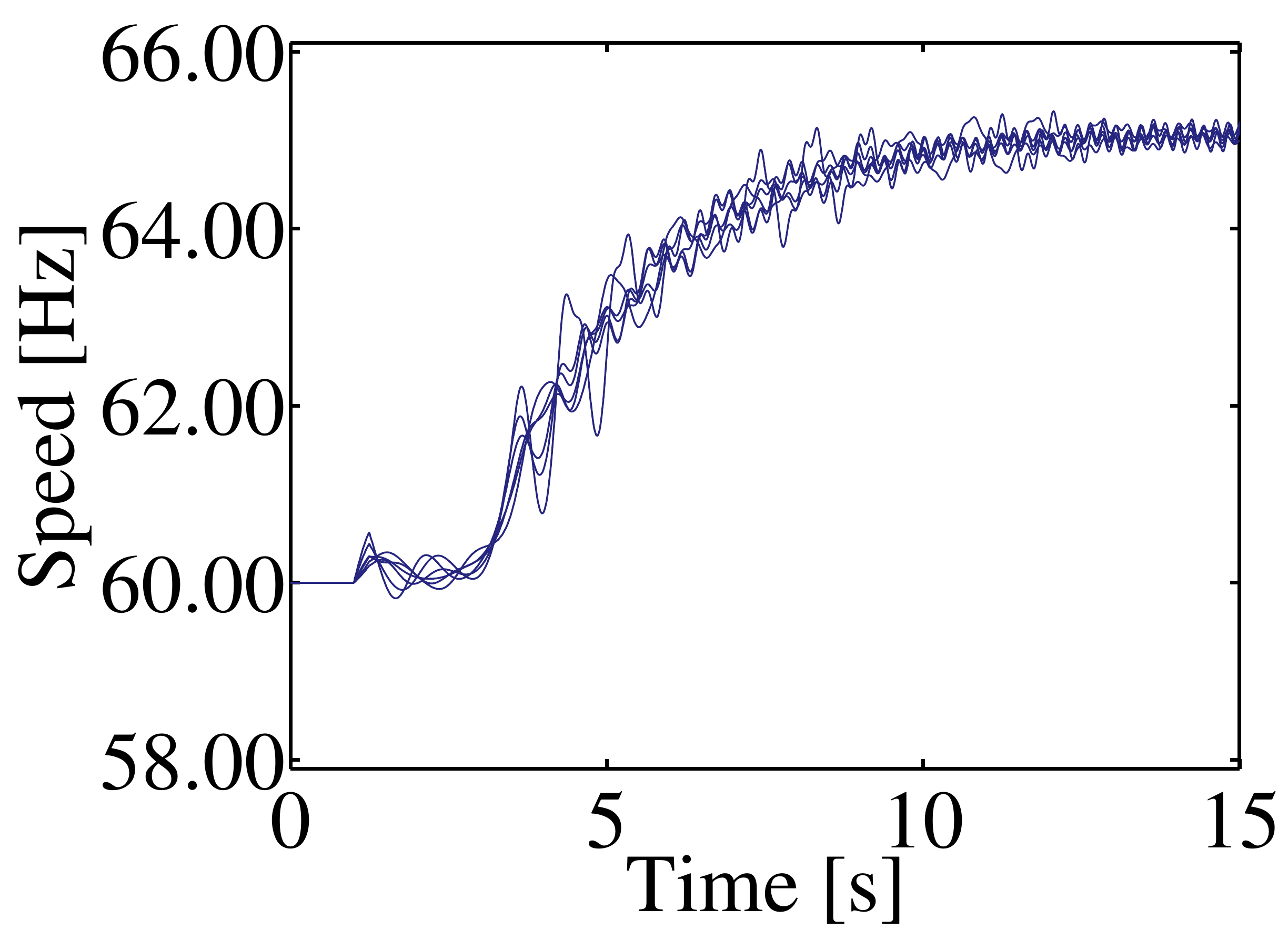}}
&
\captionsetup[subfigure]{margin={0.7cm,0cm}}
\subfloat[]{\includegraphics[scale=.2]{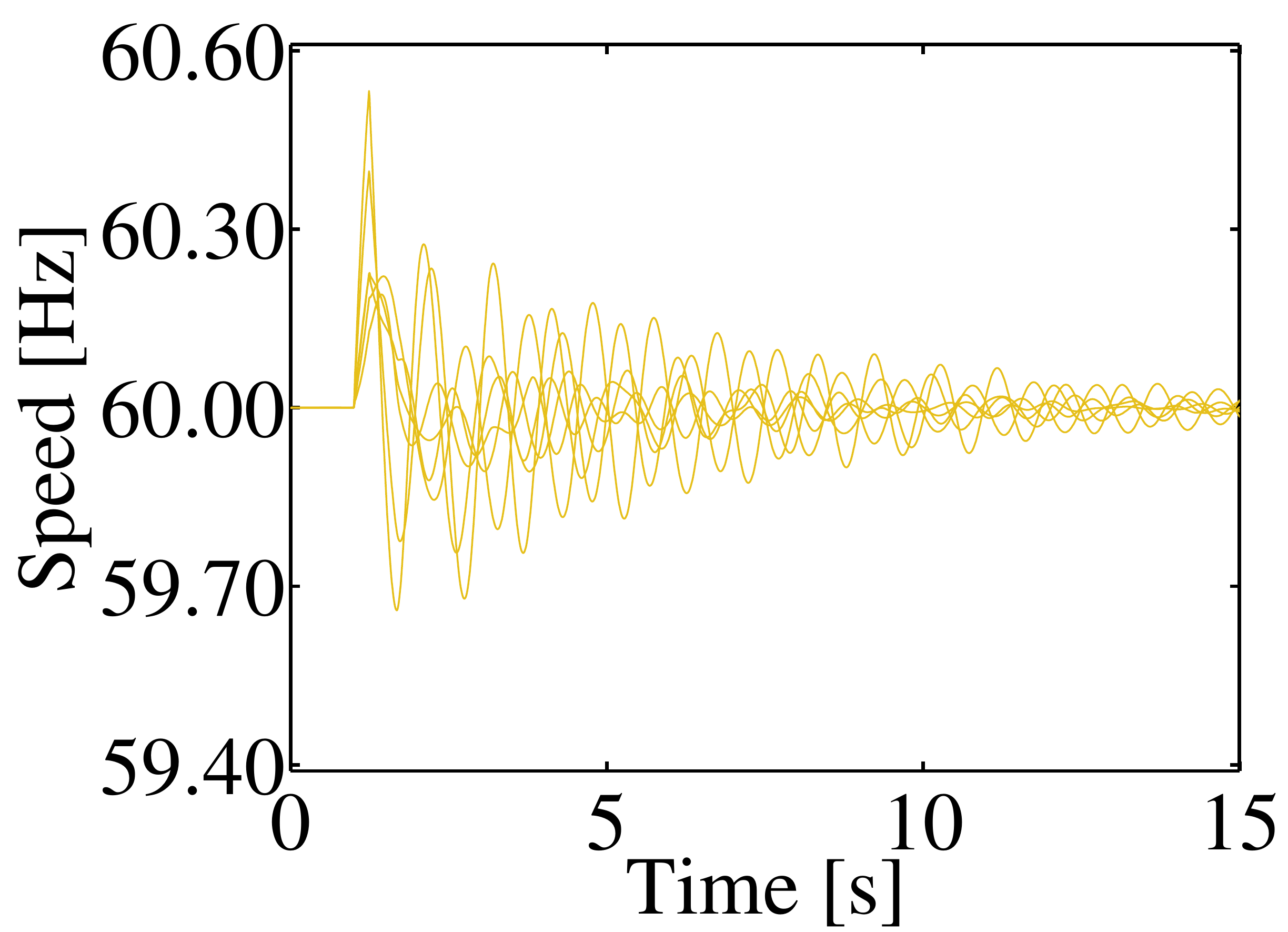}}
\end{tabular}
\caption{Turbine speeds during a bus fault at bus 12 that starts at the 1 second mark and lasts for 0.23 seconds ($t_{ccl} = 0.2276$ seconds without ACVGs). The line colors correspond to deviation from the nominal value of 60 Hz. Parameters that deviate strongly are colored purple(dark gray), while those with minor deviations are colored ye;llow(light gray). Without ACVGs  the system fails to resume synchronous operation (a) as opposed to the situation when ACVGs are used (b).}  
\label{fig:extended_tccl}
\end{figure}

To demonstrate the importance of extended critical clearing time, we simulate a bus fault at bus 12. For this disturbance $t_{ccl} = 0.2276$ seconds, which increases to $0.2715$ seconds when ACVGs are used. If the bus fault lasts longer than 0.2276 seconds, the system without ACVGs loses stability (Fig.~\ref{fig:extended_tccl}). Without ACVGs, frequency deviations go beyond the  generally adopted safety limits of 1 [Hz] (Fig.~\ref{fig:extended_tccl}.a) and voltage values continuously vary from 0-1 [PU] (Fig.~\ref{fig:extended_tccl}.c), which indicates a complete loss of synchronous operation. With ACVGs the system manages to stabilize after going through severe perturbations in both frequency and voltage (Fig.~\ref{fig:extended_tccl}.b and Fig.~\ref{fig:extended_tccl}.d).

The presented results are based on 1 percent PEV penetration as detailed in Sect.~\ref{sec:simulations_method}.
We investigated what will happen when the  penetration of plug-in electric vehicles reaches 2, 5 and even 10 percent.
We use the average increase in critical clearing times for all possible bus faults to compare the different PEV penetration scenarios. 

\begin{figure}[!htb]
\centering
\includegraphics[scale=.45]{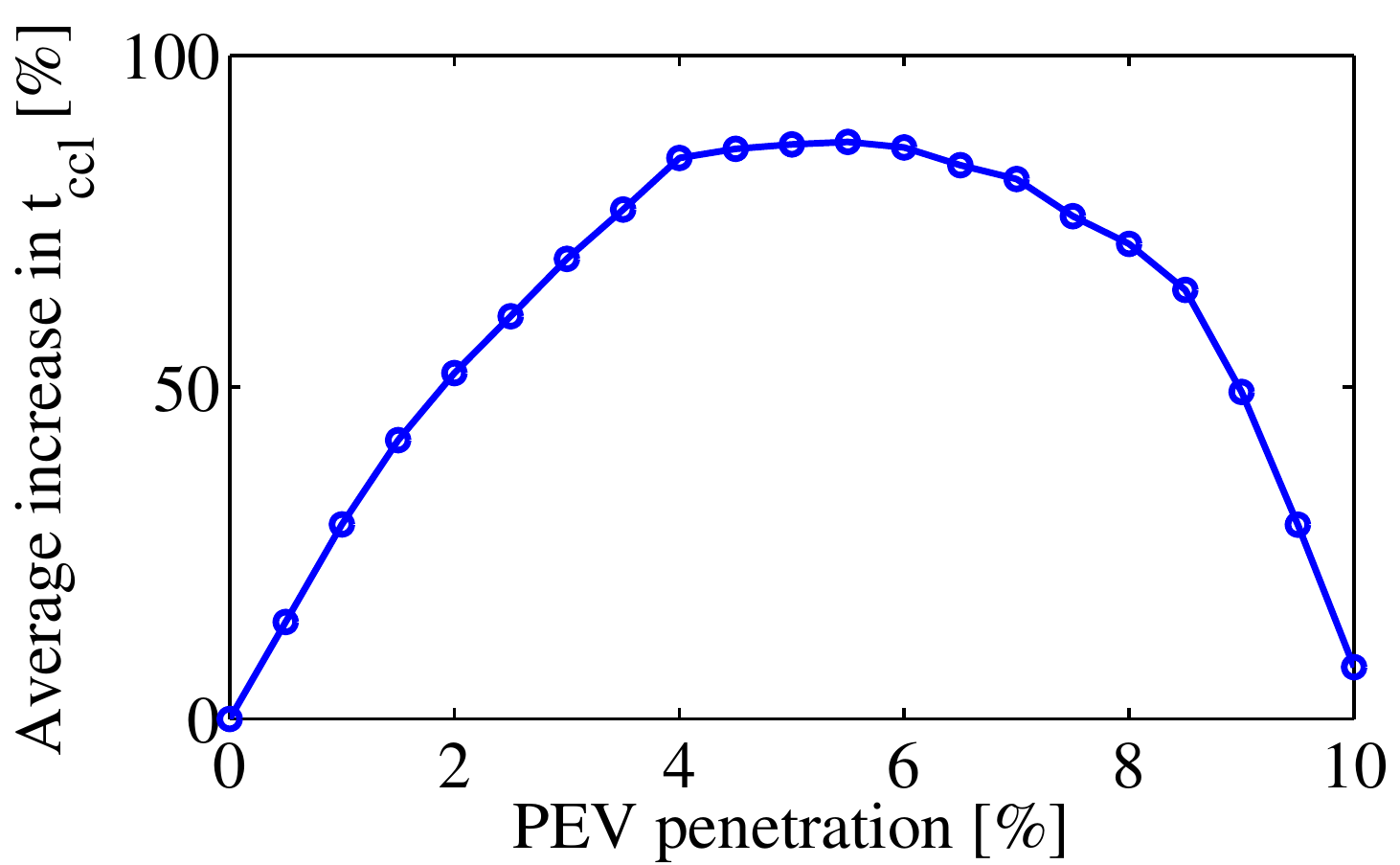}
\caption{The stability benefit of additional PEVs increases steadily up to 4\%, but degrades rapidly after 7\% penetration is reached. The average increase in the critical clearing times is calculated with respect to the original New England power system, and serves as an indicator to the effectiveness of a certain PEV penetration scenario. }
\label{fig:npev_vs_tccl}
\end{figure}

Surprisingly, the simulations show that the benefits of extra vehicles peaks at 5.5 percent penetration (Fig.~\ref{fig:npev_vs_tccl}).
This is due to the ACVGs overreacting to the disturbance and injecting or withdrawing too much power from the grid, that do not serve to stabilize the grid.
This experiment suggests that there is a limit to how much improvement can be achieved with this type of control. When this limit is reached connecting additional PEVs to serve as grid regulators has a negative impact on grid stability. The overreaction of ACVGs is directly dependent on the value of constants $h_i$. This indicates that further investigation may be directed in the area of their optimal selection which will prevent over or underreaction of ACVGs.

\section{Conclusion}

In this paper we have shown that PEVs can substantially improve power grid transient stability by designing a control strategy that regulates the power output of a fleet of PEVs based on the speed of generator turbines. The control strategy has been tested on the New England 10-unit 39-bus power system. By regulating the power output of the PEVs we have shown that (1) speed and voltage fluctuations following a large disturbance can be significantly reduced up to 80\%, and (2) the critical clearing time can be extended by 20-40\%. Future work will focus on a more general control strategy that can cope with different PEV penetration scenarios.

Although there are many benefits of V2G systems, including the improved transient stability suggested in this paper, the V2G concept faces several challenges that should be addressed before the V2G technology becomes widely accepted. The increased number of PEVs may impact power distribution system dynamics and performance through overloading of transformers, cables, and feeders.
Other barriers to the V2G transition include investment cost, resistance of automotive and oil sectors, and customer acceptance.
Two biggest challenges are battery technology (PEV batteries should have an extended cycle life, use lower cost and lightweight materials, and be more efficient) and fast and reliable two-way secure communication infrastructure network, which is needed to enable V2G technology. Nevertheless, we believe that the V2G concept has bright future as two large energy conversion systems, namely the electric utility system and the light vehicle fleet, which initially were developed separately will merge at some level in this century. 

\section*{Acknowledgment}
\addcontentsline{toc}{section}{Acknowledgment}

JK is supported by the Government of the Russian Federation (Agreement No. 14.Z50.31.0033) and IRTG 1740 (Deutsche Forschungsgemeinschaft).


\end{document}